# Models to Reduce the Complexity of Simulating a Quantum Computer

Kevin M. Obenland and Alvin M. Despain

*Information Sciences Institute / University of Southern California*

## Abstract

*Recently Quantum Computation has generated a lot of interest due to the discovery of a quantum algorithm which can factor large numbers in polynomial time. The usefulness of a quantum computer is limited by the effect of errors. Simulation is a useful tool for determining the feasibility of quantum computers in the presence of errors. The size of a quantum computer that can be simulated is small because faithfully modeling a quantum computer requires an exponential amount of storage and number of operations. In this paper we define simulation models to study the feasibility of quantum computers. The most detailed of these models is based directly on a proposed implementation. We also define less detailed models which are exponentially less complex but still produce accurate results. Finally we show that the two different types of errors, decoherence and inaccuracies, are uncorrelated. This decreases the number of simulations which must be performed.*

# 1.0 Introduction

A quantum computer consists of atomic particles which obey the laws of quantum mechanics[LaTu95][TuHo95][Lloy95]. The complexity of a quantum system is exponential with respect to the number of particles. Performing computation using these quantum particles results in an exponential amount of calculation in a polynomial amount of space and time [Feyn85][Beni82][Deut85]. This *quantum parallelism* is only applicable in a limited domain. Prime factorization is one such problem which can make effective use of quantum parallelism[Shor94]. This is an important problem because the security of the RSA public-key cryptosystem relies on the fact that prime factorization is computationally difficult[RiSA78].

Errors limit the effectiveness of any physical realization of a quantum computer. These errors can accumulate over time and render the calculation useless[ObDe96a]. The simulation of quantum circuits is a useful tool for studying the feasibility of quantum computers [ObDe96b]. Simulations inject errors at each step of the calculation and can track their accumulation.

Because of the exponential behavior of quantum computers, simulating them on conventional computers requires an exponential amount of operations and storage. Simulation of even a small problem requires a large amount of memory and a long simulation time. In this paper we define simulation models which reduce the overall complexity of simulating a quantum computer, and we show that these models are very accurate compared to the more detailed models. These simplified models represent an exponential decrease in processing and memory requirements and allow us to simulate problems which are not feasible using the more detailed models.

The rest of this paper is organized as follows. Section 2 introduces quantum computers and discusses the physical model used as the basis of our simulations. In Section 3.0 we describe our simulation methodology, and define each of the error models. Section 3.0 also describes the two types of circuits used in our simulation studies, circuits for quantum factoring and a circuit for Grover's database search problem. Section 4.0 shows the results of our simulations, which validate the reduced complexity simulation models and other simplifying assumptions. Finally in Section 5.0 we present our conclusions.

# 2.0 Quantum computers

A quantum computer performs operations on bits, called *qubits*, whose values can take on the value of one or zero or a superposition of one and zero. This superposition allows the representation of an exponential number of states using a polynomial number of qubits. A quantum computer performs transformations on these qubits to implement logic gates. Combinations of these logic gates define quantum circuits.

## 2.1 Qubits and quantum superposition

The basic unit of storage in a Quantum Computer is the *qubit*. A qubit is like a classical bit in that it can be in two states, zero or one. The qubit differs from the classical bit in that, because of the properties of quantum mechanics, it can be in both these states simultaneously[FeLS65]. A qubit which contains both the zero and one values is said to be in the *superposition* of the zero and one states. The superposition state persists until we perform an external measurement. This measurement operation forces the state to one of the two values. Because the measurement determines without doubt the value of the qubit, we must describe the possible states which exist before the measurement in terms of their probability of occurrence. These qubit probabilities must always sum to one because they represent all possible values for the qubit.

| Bit Value | Amplitude | Probability |
|---|---|---|
| 0 | 1 | 1 |
| 1 | 0 | 0 |

**(a)** Representation of a 0 qubit value

| Bit Value | Amplitude | Probability |
|---|---|---|
| 0 | $1/(\sqrt{2})$ | $1/2$ |
| 1 | $1/(\sqrt{2})$ | $1/2$ |

**(b)** Representation of a superposition between 0 and 1

**FIGURE 1. Vector representation of qubit values**

The quantum simulator represents a qubit using a complex vector space. Figure 1 shows how the simulator uses complex amplitudes to represent a qubit. Each state in the vector represents one of the possible values for the qubit. The bit value of a state corresponds to the index of that state in the vector. The simulator represents each encoded bit value with a non zero amplitude in the state vector. The probability of each state is defined as the square of this complex amplitude[FeLS65]. Figure 1(a) shows a state which represents the single value of zero. In Figure 1(b) the probability is equally split between the zero and one states, representing a qubit which is in the superposition state. For a register with M qubits, the simulator uses a vector space of dimension $2^M$.

An M qubit register can represent $2^M$ simultaneous values by putting each of the bits into the superposition state. A calculation using this register calculates all possible outcomes for the $2^M$ input values, thereby giving exponential parallelism. The bad news is that in order to read out the results of a calculation we have to observe, i.e. measure, the output. This measurement forces all the qubits to a particular value thereby destroying the parallel state. The challenge then is to devise a quantum calculation where we can accumulate the parallel state in non-exponential time before performing the measurement.

## 2.2 Quantum transformations and logic gates

A quantum computation is a sequence of transformations performed on the qubits contained in quantum registers[Toff81][FrTo82][Feyn85][BaBe95][Divi95]. A transformation takes an input quantum state and produces a modified output quantum state. Typically we define transformations at the gate level, i.e. transformations which perform logic functions. The simulator performs each transformation by multiplying the $2^M$ dimensional vector by a $2^M$ x $2^M$ transformation matrix.

The basic gate used in quantum computation is the *controlled-not*, i.e. exclusive or gate. The controlled-not gate is a two bit operation between a *control* bit and a *resultant* bit. The operation of the gate leaves the control bit unchanged, but conditionally flips the resultant bit based on the value of the control bit. Table 1 shows a truth table of how the controlled-not gate modifies the different qubit values. In the vector representation of the qubits, the controlled-not gate corresponds to a transformation which swaps the amplitude of the states in the third and fourth positions. Figure 2 shows the four by four matrix which performs the controlled-not transformation on the two qubits.

**TABLE 1. Truth table for the controlled-not gate**

| Input Bits | | Output Bits | |
|---|---|---|---|
| A | B | A' | B' |
| 0 | 0 | 0 | 0 |
| 0 | 1 | 0 | 1 |
| 1 | 0 | 1 | 1 |
| 1 | 1 | 1 | 0 |

$$\begin{bmatrix} 1 & 0 & 0 & 0 \\ 0 & 1 & 0 & 0 \\ 0 & 0 & 0 & 1 \\ 0 & 0 & 1 & 0 \end{bmatrix} \times \begin{bmatrix} a_0 \\ a_1 \\ a_2 \\ a_3 \end{bmatrix} = \begin{bmatrix} a_0 \\ a_1 \\ a_3 \\ a_2 \end{bmatrix} \longleftarrow \textbf{Swap amplitudes } a_2 \textbf{ and } a_3$$

**FIGURE 2. Controlled not transformation**

## 2.3 The trapped ion quantum computer

The ion trap quantum computer as proposed by Cirac and Zoller is one of the most promising schemes for the experimental realization of a quantum computer[CiZo95]. To date cavities trapping up to 33 ions have been constructed[RaGi92], and simple quantum gates have been demonstrated[MoMe95][WiMM96]. Laser pulses directed at the ions in the trap cause transformations to their internal state. A controlled-not gate is a sequence of these laser pulses. The trapped ion quantum computer model is the basis of all the simulation models described in this paper.

### 2.3.1 Qubits in the ion trap quantum computer

Qubits are represented using the internal energy states of the ions in the trap. The ion trap represents a logic zero with the ground state of an ion, and a logic one with a higher energy state. The ion trap quantum computer also requires a third state which it uses to implement the controlled-not gate. This third state adds to the processing and memory requirements needed to simulate the quantum computer. To represent M qubits, we now require $3^M$ states, and transformations are now matrix multiplications using matrices of size $3^M \times 3^M$. In this paper we show that a simplified model of the ion trap, which uses only two states to represent a qubit, is very accurate as compared to the detailed model which uses all three states.

The ion trap uses a common phonon vibration mode to communicate between qubits in the trap. A laser pulse directed at one qubit excites the phonon mode. This phonon excitation affects the operation of a subsequent laser operation on another qubit. Representation of the phonon mode requires an extra state for each qubit, doubling the memory requirements.

Figure 3 shows the implementation of a single qubit in the ion trap. The two states $|g\rangle$ and $|e_0\rangle$ represent the two qubit values zero and one. The state $|e_1\rangle$ serves as the auxiliary state. Laser pulses directed at the ion cause it to transition between states. A laser pulse can also transfer a state to and from the phonon vibration mode. By tuning the laser frequency we can cause transitions between different states. If the ion is in a state other than the one the pulse is tuned to, it is unaffected by the laser pulse. Each laser pulse of duration π, multiplies the amplitude by a factor of -i, and therefore two identical transformations, i.e. a 2π pulse, will return the ion to the original state but flip the sign of the amplitude.

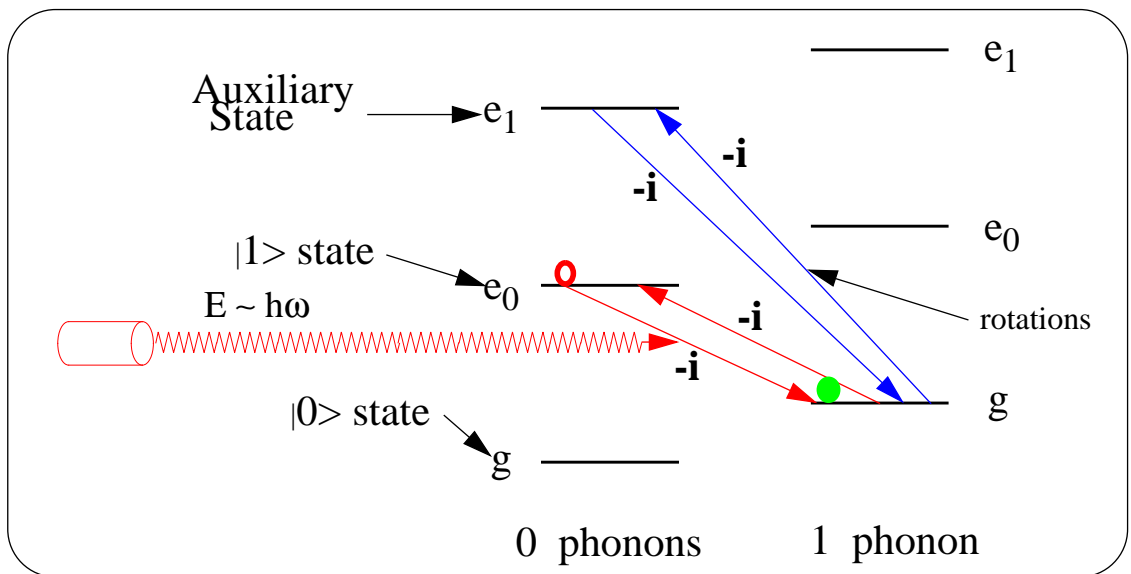

**FIGURE 3. Laser transformations on an ion and a phonon**

### 2.3.2 Transformations in the ion trap

An operation in the ion trap quantum computer is a sequence of laser pulses. Each laser pulse is defined by one of the transformation matrices shown in Figure 4. We can implement various transformations by selecting different values for the angles θ and ϕ. θ corresponds to the duration of the laser pulse and ϕ corresponds to the phase. Both the U and V transformations are single bit rotations. The V transformation operates only on the qubit, and the U transformation operates on the qubit as well as the common phonon mode.

Each laser pulse performs a rotation between two states, as shown in Figure 3. Tuning the laser to different transitions results in rotations between different pairs of states. In this paper we always use the laser tuning defined by Equation 1 for the V transformation. For the U transformation, we use the three tunings shown in Equation 2 through Equation 4.

$$U(\theta,\phi) = \begin{bmatrix} 1 & 0 & 0 & 0 \\ 0 & \cos\frac{\theta}{2} & (-i)e^{-i\phi}\sin\frac{\theta}{2} & 0 \\ 0 & (-i)e^{i\phi}\sin\frac{\theta}{2} & \cos\frac{\theta}{2} & 0 \\ 0 & 0 & 0 & 1 \end{bmatrix} \qquad V(\theta,\phi) = \begin{bmatrix} \cos\frac{\theta}{2} & (-i)e^{-i\phi}\sin\frac{\theta}{2} \\ (-i)e^{i\phi}\sin\frac{\theta}{2} & \cos\frac{\theta}{2} \end{bmatrix}$$

**FIGURE 4. The U and V transformation matrices which describe all the operations in the trapped ion quantum computer**

$$V(\theta, \phi): \qquad |g\rangle \Leftrightarrow |e_0\rangle \qquad \text{(EQ 1)}$$

$$U(\theta, \phi): \qquad |e_0\rangle|0\rangle_p \Leftrightarrow |g\rangle|1\rangle_p \qquad \text{(EQ 2)}$$

$$\tilde{U}(\theta, \phi): \qquad |e_1\rangle|0\rangle_p \Leftrightarrow |g\rangle|1\rangle_p \qquad \text{(EQ 3)}$$

$$\hat{U}(\theta, \phi): \qquad |e_1\rangle|0\rangle_p \Leftrightarrow |e_0\rangle|1\rangle_p \qquad \text{(EQ 4)}$$

### 2.3.3 Implementing the controlled-not gate

The controlled-not gate operates on two qubits and the phonon. The first qubit, the *control* qubit, conditionally flips the second qubit, the *resultant* qubit. A gate operation always starts and ends in the |0> phonon state. In the ion trap the controlled-not gate is a sequence of five laser pulses, one of which is the $U_m$ transformation shown in Equation 5, which operations on the control qubit $m$.

$$U_m(\pi, 0) = \begin{bmatrix} 1 & 0 & 0 & 0 \\ 0 & 0 & -i & 0 \\ 0 & -i & 0 & 0 \\ 0 & 0 & 0 & 1 \end{bmatrix} \qquad \text{(EQ 5)}$$

$$\tilde{U}_n(2\pi, 0) = \begin{bmatrix} 1 & 0 & 0 & 0 \\ 0 & -1 & 0 & 0 \\ 0 & 0 & 1 & 0 \\ 0 & 0 & 0 & 1 \end{bmatrix} \qquad \text{(EQ 6)}$$

The $\tilde{U}_n(2\pi, 0)$ transformation, shown in Equation 6, performs a rotation of the resultant qubit through the auxiliary state. This transformation leaves the resultant bit in the same state and changes the sign of the $|g\rangle_n|1\rangle_p$ state.

The sequence $U_m(\pi, 0) \bullet \tilde{U}_n(2\pi, 0) \bullet U_m(\pi, 0)$, as shown in Equation 7, implements a controlled phase gate. We can use this controlled phase gate to implement a controlled-not gate(see below). The first $U_m$ transformation transfers the state of the control bit $m$ to the phonon mode. The $\tilde{U}_n$ transformation then rotates the state $|g\rangle_n|1\rangle_p$ through the auxiliary state. This leaves the resultant bit in the same state but changes the sign of the amplitude. Applying another $U_m$ removes the phonon state. The net effect of these three transformations is to negate the sign of the amplitude only when both qubits $n$ and $m$ are in the excited state, i.e. the state $|e_o\rangle_m|e_o\rangle_n$. This negation of the sign of the amplitude corresponds to a phase change of $180^o$.

$$
\begin{array}{cccc}
& U_m(\pi, 0) & \tilde{U}_n(2\pi, 0) & U_m(\pi, 0) \\
|g\rangle_m|g\rangle_n|0\rangle_p & |g\rangle_m|g\rangle_n|0\rangle_p & |g\rangle_m|g\rangle_n|0\rangle_p & |g\rangle_m|g\rangle_n|0\rangle_p \\
|g\rangle_m|e_0\rangle_n|0\rangle_p & |g\rangle_m|e_0\rangle_n|0\rangle_p & |g\rangle_m|e_0\rangle_n|0\rangle_p & |g\rangle_m|e_0\rangle_n|0\rangle_p \\
|e_0\rangle_m|g\rangle_n|0\rangle_p \Rightarrow & -i|g\rangle_m|g\rangle_n|1\rangle_p \Rightarrow & i|g\rangle_m|g\rangle_n|1\rangle_p \Rightarrow & |e_0\rangle_m|g\rangle_n|0\rangle_p \\
|e_0\rangle_m|e_0\rangle_n|0\rangle_p & -i|g\rangle_m|e_0\rangle_n|1\rangle_p & -i|g\rangle_m|e_0\rangle_n|1\rangle_p & -|e_0\rangle_m|e_0\rangle_n|0\rangle_p
\end{array}
\qquad \text{(EQ 7)}
$$

To get the controlled-not operation from the $U_m(\pi, 0) \bullet \tilde{U}_n(2\pi, 0) \bullet U_m(\pi, 0)$ transformations we apply the transformation $V_n(\pi/2, -\pi/2)$ to the resultant bit before the three transformations, and apply the inverse operation $V_n(\pi/2, \pi/2)$ after the three transformations. The matrices which describe the two $V_n$ transformations are shown in Equation 9. The first $V_n$ transformation changes the basis, i.e. the representation, of the |0> and |1> states. The |0> state is represented by $(|g\rangle - |e_0\rangle)$ in the new basis, and the |1> state is represented by $(|g\rangle + |e_0\rangle)$. These states differ only in the sign of the $|e_0\rangle$ state. Therefore if the $m$ qubit is set, the controlled phase gate will change the sign of the $|e_0\rangle$ state in the new basis, corresponding to a bit flip of the qubit $n$ in the original basis. If the $m$ qubit is not set, the controlled phase gate has no effect on the state of the qubits. The complete controlled-not gate is performed using the sequence shown in Equation 8, where the operations are applied from right to left.

$$V_n(\pi/2, \pi/2) \bullet U_m(\pi, 0) \bullet \tilde{U}_n(2\pi, 0) \bullet U_m(\pi, 0) \bullet V_n(\pi/2, -\pi/2) \qquad \text{(EQ 8)}$$

$$V_n(\pi/2, -\pi/2) = \frac{1}{\sqrt{2}}\begin{bmatrix} 1 & 1 \\ -1 & 1 \end{bmatrix} \qquad V_n(\pi/2, \pi/2) = \frac{1}{\sqrt{2}}\begin{bmatrix} 1 & -1 \\ 1 & 1 \end{bmatrix} \qquad \text{(EQ 9)}$$

## 2.4 Errors in the operation of a quantum computer

It is impossible to perform the laser operations perfectly. The resulting inaccuracies, referred to as *operational* errors, degrade the calculation over time. These inaccuracies add $\delta$ factors to the angles $\theta$ and $\phi$ in the $U$ and $V$ transformations.

Interaction of the phonon state with the external environment has a destructive effect on the coherence of the superposition state. This type of error, referred to as *decoherence*, can be thought of as a measurement of the phonon state. Decoherence destroys the parallel state turning the quantum computation into a classical one.

Both operational and decoherence errors limit the effectiveness of a quantum computer. Simulation is an effective tool for characterizing errors, and tracking their accumulation. Using a physical model such as the Cirac and Zoller scheme is important for obtaining realistic results.

## 2.5 Quantum algorithms

Much of the current interest in quantum computation is due to the discovery of an efficient algorithm to factor numbers[Shor94]. This is an important problem because a quantum factoring engine would severely threaten the security of public-key cryptosystems. The quantum factoring algorithm uses quantum parallelism to calculate all of the values of a function simultaneously. This function is periodic and a quantum FFT can extract this period efficiently [Copp94]. We then use a polynomial time classical algorithm to calculate the factors from this period.

Quantum computers are also useful for searching unsorted databases[Grov96]. The quantum search algorithm runs in $O(\sqrt{N})$ time for $N$ items, where the best classical algorithm runs in $O(N)$ time. Therefore for NP-complete search problems such as circuit SAT, which contains $2^m$ items for $m$ variables, the quantum algorithm runs in $O(2^{m/2})$ time. This speedup is not exponential like that of the factoring algorithm, but it allows the solution of problems that may be computationally intractable on a classical computer.

# 3.0 Quantum simulation

Our quantum simulator simulates circuits at the gate level. The simulator implements one, two and three bit controlled-not gates as well as rotation gates. The simulator implements each gate as a sequence of laser pulses, and represents the entire Hilbert space through out the calculation. The size of the Hilbert space depends on the level of detail of the simulation model.

## 3.1 Simulation of quantum logic gates

The simulator implements each gate as a sequence of transformations on the complex Hilbert space. In general, if our quantum computer has a Hilbert space of size V, then each transformation is a VxV matrix multiplication. Because each transformation operates on a single bit and possibly the phonon mode, these operation are simply 2x2 or 4x4 matrices replicated many times. We can avoid having to represent the entire VxV matrix by iterating over the Hilbert space and performing the simpler operations for all of the corresponding sets of states[ObDe98].

### 3.1.1 Operational errors

The simulator injects operational errors at each step by adding a small deviation to the two angles of rotation θ and φ. Each error angle is drawn from a gaussian distribution with a parametrized mean (μ) and standard deviation (σ). We vary the level of error in one of the following ways:

- **Bias errors**: Positive errors with a constant mean (μ) and a standard deviation (σ) of zero. All transformations use the same error angle throughout the simulation.

- **Noise**: Random Gaussian errors with a given standard deviation (σ) and a mean (μ) of zero. New random error angles are generated for each transformation.

- **Bias errors and noise**: Random Gaussian errors with a given standard deviation (σ) and a fixed mean (μ). This error model is a generalization of the first two error models.

Bias errors correspond to systematic calibration errors, and random gaussian errors model noise in the operation of the laser apparatus.

### 3.1.2 Decoherence errors

Because the phonon mode is coupled to all the qubits in the computer, it is the largest source of decoherence[MoMe95]. For this reason we only model the phonon decoherence and not the decoherence of the individual qubits.

We model the decoherence of the phonon mode by performing an additional operation after each laser pulse. Equation 10 shows this transformation which has the effect of decaying the amplitude of the states in the $|1\rangle_p$ phonon state. This decay transformation is based on the quantum jump method of decay[Carm93][Garg96]. The decay parameter (*dec*) remains constant throughout the entire simulation.

$$\begin{array}{l} |\psi\rangle|0\rangle_p \\ |\psi\rangle|1\rangle_p \end{array} \Rightarrow \begin{array}{l} |\psi\rangle|0\rangle_p \\ e^{(-dec)/2}|\psi\rangle|1\rangle_p \end{array} \quad \text{(EQ 10)}$$

The simulator uses two different methods for modeling decoherence. For the first method, called *spon_emit*, the simulator performs a decay transformation after every laser pulse and then checks for the occurrence of a spontaneous emission. The probability that an emission occurs, as given in Equation 11, is based on the amount of phonon amplitude present and the decoherence rate. An emission occurs if a randomly generated number is less than the emission probability. Because of this random event, we must average the results of multiple trials each starting with different initial random seeds. An emission destroys the superposition of the $|0\rangle_p$ and $|1\rangle_p$ phonon states and the amplitude from the $|1\rangle_p$ state collapses into the $|0\rangle_p$ state. The simulator renormalizes the state after a decay step or an emission, so that the total norm is always one.

$$prob_{emission} = dec \bullet prob_{phonon} \quad \text{(EQ 11)}$$

The second method for modeling decoherence, called *decay*, does not explicitly model spontaneous emission. After each laser pulse the simulator performs the decay transformation but the qubit state is not renormalized. This has the effect of decaying the total norm towards zero. The total norm at each step now represents the probability that the calculation survives up to that point without a spontaneous emission occurring. We show in Section 4.3.1 that both types of simulations are essential the same.

## 3.2 Circuits for factoring

Much of the current interest in quantum computation is due to the discovery of an efficient algorithm by Peter Shor to factor numbers[Shor94]. By putting the qubit register *A* in the superposition of all values and calculating the function $f(A) = X^A \bmod N$, a quantum computer calculates all the values of *f(A)* simultaneously. Where *N* is the number to be factored, and *X* is a randomly selected number which is relatively prime to *N*. Figure 5 shows a circuit for factoring the number 15, where we use rotation gates to prepare the superposition state. In this circuit if we use a value of 7 for *X*, we see the repeating pattern for *f(A)* shown in Table 2. To extract the period of this function we perform a quantum FFT on the amplitudes in the *A* register[Copp94]. This produces a number which contains the period, and can be used by a classical computer to obtain the factors of the number *N*.

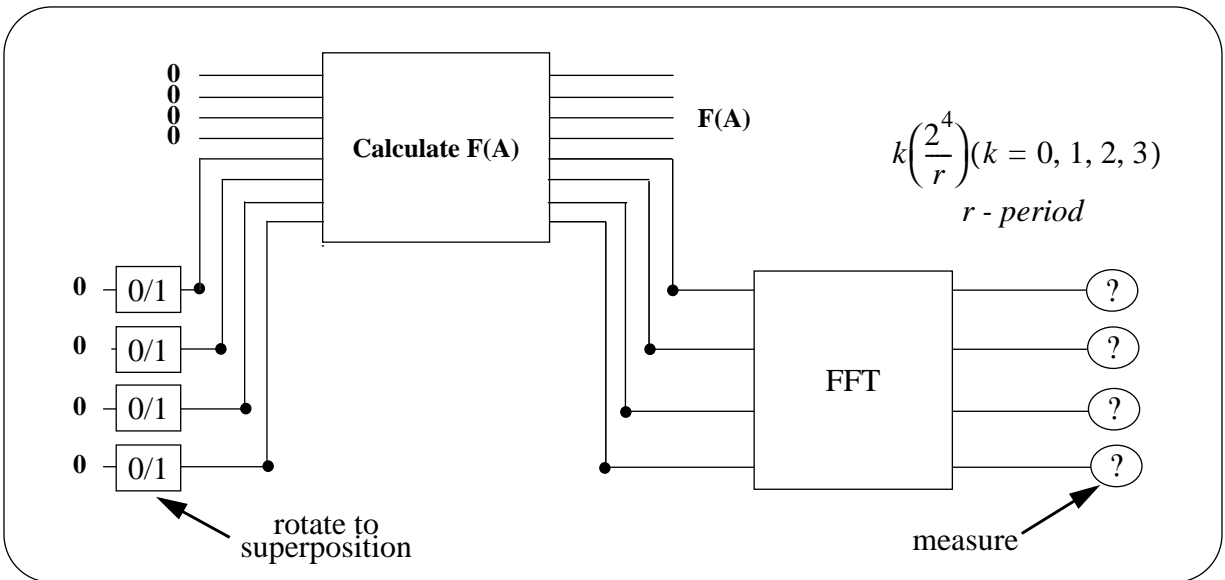

**FIGURE 5. Factor 15 Circuit used in simulations**

**TABLE 2. Function f(A) = $7^A$ mod 15**

| A | f(A) |
|---|------|
| 0 | 1 |
| 1 | 7 |
| 2 | 4 |
| 3 | 13 |
| 4 | 1 |
| 5 | 7 |
| 6 | 4 |
| 7 | 13 |
| : | : |

As we see from Figure 5 each quantum factoring circuit consists of three pieces: (1) preparing the superposition state, (2) calculating the function *f(A)* and (3) performing the quantum FFT. We do not include the portion of the calculation which extracts the period because it is performed off-line on a classical computer. The circuit to calculate *f(A)* constitutes the largest portion of the circuit and can be performed in O($L^3$) time using repeated squaring. Because we can also perform the extra processing need to extract the factor in polynomial time, Shor's algorithm gives an efficient means of factoring large numbers

### 3.2.1 Factor 21 circuit using table lookup

One method for calculating the *f(A)* function is to perform a direct mapping from the input values (*A*) to the output, i.e. *f(A)*[ObDe96a]. We start with all bits in the *f(A)* register in the zero state. We then use NOT gates to select each possible value for *A*, and flip the bits of *f(A)* to the correct value. This method is only valid for factoring small numbers because it is exponential in the number of input bits. This table lookup methods allows the detailed simulation of relatively small circuits. Factoring the number 21 using this method requires about 400 gates.

### 3.2.2 General factoring

We use repeated squaring to perform the modulo exponentiation required by the *f(A)* circuit[Desp96][BeCh96]. Equation 12 shows that using the binary representation of *A*, the exponentiation consists of a sequence of multiplications. At each step we use a single bit of A and multiply a running product by the factor $X^{2^l}$. Each of these multiplications requires time $O(L^2)$, where $L = \log_2 N$. In order to extract the period from the quantum calculation, Shor suggests using 2L + 1 bits for the input register A[Shor94].

$$f(A) = X^A mod\ N = X^{[a_0 2^0 + a_1 2^1 ... a_{l-1} 2^{l-1}]} mod\ N = X^{a_0 2^0} \bullet X^{a_1 2^1} ... X^{a_{l-1} 2^{l-1}} mod\ N \quad \text{(EQ 12)}$$

The circuit used in this paper also requires two additional scratch registers to write the intermediate results of addition and multiplication. The number of qubits required to factor an *L* bit number using this circuit is therefore, *5L + 4*. This assumes a maximum fan-in of three. The total number of operations using this method is $252L^3 + 8L^2 + L + 3$.

## 3.3 Database search in the ion trap

Grover's database search algorithm uses a mixture of logic gates, i.e. gates such as the controlled-not gate, and rotations. We implement the logic gates as shown in Section 2.3.3. In this section we review Grover's Database search algorithm and describe a method to implement it in the ion trap quantum computer.

### 3.3.1 Transformations needed for the Grover algorithm

The database search starts by initializing the qubit state to the superposition state shown in Equation 13. This state defines a search space of N items using $log_2 N$ qubits.

$$|\psi_0\rangle = \frac{1}{\sqrt{N}} \sum_{l=0}^{N-1} |l\rangle \qquad \text{(EQ 13)}$$

The next step is to evaluate the function $f(l)$ and write the result into the temporary qubit state $|r\rangle$. The function $f(l)$ is the search function which identifies the matching keys in the database search. $f(l)$ evaluates to one if a state value matches the key and zero otherwise. In general there can be multiple matching keys. In the circuit SAT problem $f(l)$ is just one evaluation of the logic function which describes the circuit.

The next step is to rotate the phase of the states where $|r\rangle$ is one by $\pi$ radians. This has the effect of negating the sign of the states with matching key values. The obvious way to evaluate $f(l)$ and perform the rotation is to start with $|r\rangle$ in the $|0\rangle$ state, evaluate $f(l)$ and then perform a conditional rotation for the resulting states which have $|r\rangle$ equal to $|1\rangle$. This requires an additional application of $f(l)$ to return $|r\rangle$ to $|0\rangle$. An alternative approach it to start with $|r\rangle$ in the state $(|1\rangle - |0\rangle)$. This combines the evaluation of $f(l)$ with the rotation and requires only a single application of $f(l)$ [BaBe95]. After this step the original state $|\psi_0\rangle$ is transformed into the state $|\psi_1\rangle$ as shown in Equation 14.

$$|\psi_1\rangle = C|\psi_0\rangle = \frac{1}{\sqrt{2N}} \sum_{l=0}^{N-1} (-1)^{f(l)} |l\rangle (|1\rangle - |0\rangle) \qquad \text{(EQ 14)}$$

The next step is to apply the diffusion transformation D. Equation 15 shows an example of a diffusion matrix for N=4. The diagonal terms have values $(2/N)$ and all other terms have values $((2/N) - 1)$.

$$D_4 = \begin{bmatrix} -1/2 & 1/2 & 1/2 & 1/2 \\ 1/2 & -1/2 & 1/2 & 1/2 \\ 1/2 & 1/2 & -1/2 & 1/2 \\ 1/2 & 1/2 & 1/2 & -1/2 \end{bmatrix} \qquad \text{(EQ 15)}$$

A convenient method of implementing the diffusion matrix is to use the sequence of transformations *FRF*. *F*, as shown in Equation 16, is the single bit Fourier transformation matrix[BaEk96]. To perform the *F* transformation on the qubit register $|l\rangle$ we apply the two by two transformation $F_2$ to each of the bits in $|l\rangle$. R is an NxN matrix transformation which negates the sign of all the states in the $|l\rangle$ register except for the state $|0...0\rangle$. An example of R for N=4 is also shown in Equation 16.

$$F_2 = \frac{1}{\sqrt{2}}\begin{bmatrix} 1 & 1 \\ 1 & -1 \end{bmatrix} \qquad R_4 = \begin{bmatrix} 1 & 0 & 0 & 0 \\ 0 & -1 & 0 & 0 \\ 0 & 0 & -1 & 0 \\ 0 & 0 & 0 & -1 \end{bmatrix} \qquad \text{(EQ 16)}$$

A single step in the grover algorithm is the sequence of transformations *FRFC*, i.e. *DC*. This sequence is repeated until the probability of measuring a state $|l\rangle$ where $f(l) = 1$ is greater than a minimum probability. Grover shows that after $O(\sqrt{N})$ iterations the probability of measuring this state is at least $1/2$. The number of iterations must be picked carefully because the probability decreases if the algorithm is run too long[BoBr96].

Figure 6 shows the result of a single iteration of the grover algorithm for N=4. The initial state $|\psi\rangle$ starts in an equal superposition of N values. For simplification Figure 6 does not show the state of the $|r\rangle$ register. In the example $f(0) = 1$, and for all other values of $l$ $f(0) = 0$. This produces a matrix which is the same as the matrix D except that the sign of the elements in the first column are negated.

The diffusion matrix, when applied to the state $|\psi\rangle$, amplifies the amplitude in the state $|00\rangle$ because all the elements in the first row of the matrix are positive. The diffusion matrix attenuates all other states because of the negative elements in these states corresponding rows in the matrix. For this example only a single iteration was needed, but in general we will need to perform multiple iterations.

$$DC(l=0) = \begin{bmatrix} 1/2 & 1/2 & 1/2 & 1/2 \\ -1/2 & -1/2 & 1/2 & 1/2 \\ -1/2 & 1/2 & -1/2 & 1/2 \\ -1/2 & 1/2 & 1/2 & -1/2 \end{bmatrix} \qquad |\psi\rangle = \begin{bmatrix} 1/2 \\ 1/2 \\ 1/2 \\ 1/2 \end{bmatrix} \qquad DC|\psi\rangle = \begin{bmatrix} 1 \\ 0 \\ 0 \\ 0 \end{bmatrix}$$

**FIGURE 6. One iteration of the grover database search algorithm**

### 3.3.2 The Grover algorithm in the ion trap

To implement the Grover algorithm, in the ion trap, we use a sequence of *U* and *V* matrix transformations. Figure 4 shows the format of the matrices which implement these two transformations.

To implement the Grover algorithm we need to implement the *C*, *F*, and *R* transformations as well as a general rotation to prepare the initial state. We can use the $V(\theta, \pi/2)$ transformation to implement a general rotation of the angle $\theta$. To prepare the superposition state in the $|l\rangle$ register we rotate each bit in the register by $\pi$ radians. To prepare the state $(|1\rangle - |0\rangle)$ in the $|r\rangle$ state we first set the qubit value to one, and then perform the transformation $V(\pi/2, \pi)$. To set a qubit to a the value of one, it is first measured forcing the value to one or zero. If the measured value is zero a single bit rotation is used to rotate it to the one value. Clearing a qubit is performed in the same manner.

The C transformation implements the characteristic function $f(l)$ as a sequence of one, two and three bit controlled-not gates. We implement the *F* transformation, as shown in Equation 17, using one *V* and one *U* transformation. We describe the *U* transformation using a 2x2 matrix because the phonon mode starts and ends in the $|0\rangle$ state.

$$U(2\pi, 0) \bullet V(\pi/2, -\pi) = \begin{bmatrix} 1 & 0 \\ 0 & -1 \end{bmatrix} \bullet \frac{1}{\sqrt{2}} \begin{bmatrix} 1 & 1 \\ -1 & 1 \end{bmatrix} = \frac{1}{\sqrt{2}} \begin{bmatrix} 1 & 1 \\ 1 & -1 \end{bmatrix} \quad \text{(EQ 17)}$$

The implementation of the R transformation requires the use of the extra register $|s\rangle$ which is set to a qubit value of $|1\rangle$ at the start of the computation. $|s\rangle$ is always a single bit register regardless of the size of the problem. R, as shown in Equation 18, operates on the $|l\rangle|s\rangle$ registers and requires 2L + 1 U transformations for L qubits in the $|l\rangle$ register. In Equation 18, $U_{n_j}$ denotes a U transformation acting on the qubit indexed by j.

$$R = U_{n_s}(\pi, 0) \bullet \prod_{j = L-2}^{0} \hat{U}_{n_j}(-\pi, 0) \bullet \tilde{U}_{n_L}(2\pi, 0) \bullet \prod_{j = 0}^{L-2} \hat{U}_{n_j}(\pi, 0) \bullet U_{n_s}(\pi, 0) \quad \text{(EQ 18)}$$

## 3.4 Quantum simulation models

We use several techniques to reduce the complexity of simulating a quantum computer. One technique is to reduce, from three to two, the number of states needed to represent a qubit. We can also use special properties of the factor-15 circuit to reduce its simulation complexity further.

### 3.4.1 Reducing the qubit representation to two bits

Eliminating one of the states in the representation of a qubit reduces the simulation complexity exponentially. The full 3-State model and this reduced model are defined below:

- **3-State**: Full simulation using the Cirac and Zoller trapped ion Quantum computer. Each qubit using this model requires three states to represent it for a total of $3^M * 2$ states for M qubits.
- **2-State**: This model uses only a single third state which is shared amongst all the qubits in the computer. Therefore for M qubits, we need only $2^{M+2}$ states.

The third state of a qubit in the ion trap is used only as a temporary state because all transformations through it are always performed in pairs, i.e. two $\pi$ rotations or a single $2\pi$ rotation. Equation 19 shows the use of the $\tilde{U}$ transformation in the controlled-not gate. The first $\tilde{U}$ rotates the state $|g\rangle_n$ to the $|e_1\rangle_n$ state and the second transformation rotates it back. The net result is that the phase of the state is negated.

$$\begin{array}{c} |g\rangle_n|1\rangle_p \\ |e_0\rangle_n|1\rangle_p \end{array} \xRightarrow{\tilde{U}_n} \begin{array}{c} -i|e_1\rangle_n|0\rangle_p \\ |e_0\rangle_n|1\rangle_p \end{array} \xRightarrow{\tilde{U}_n} \begin{array}{c} -|g\rangle_n|1\rangle_p \\ |e_0\rangle_n|1\rangle_p \end{array} \quad \text{(EQ 19)}$$

Equation 20 shows the two $U$ transformations used in the controlled-controlled not gate, one on the resultant bit ($n$) and one on one of the control bits ($m$). As before all amplitude is rotated out of the third state at the completion of the sequence. Because the first transformation rotates the state $|g\rangle_m|g\rangle_n|1\rangle_p$ to the $|0\rangle_p$ phonon state, the second transformation has no effect. In general it will always be the case that each state is affected by at most one of the pairs of transformations. We can use this fact to implement all consecutive operations which use the third state with a single pass over the state space.

$$
\begin{array}{ccccc}
& \tilde{U}_m(\pi, 0) & \tilde{U}_n(\pi, 0) & \tilde{U}_n(\pi, 0) & \tilde{U}_m(\pi, 0) \\
|g\rangle_m|g\rangle_n|1\rangle_p & -i|e_1\rangle_m|g\rangle_n|0\rangle_p & -i|e_1\rangle_m|g\rangle_n|0\rangle_p & -i|e_1\rangle_m|g\rangle_n|0\rangle_p & -|g\rangle_m|g\rangle_n|1\rangle_p \\
|g\rangle_m|e_0\rangle_n|1\rangle_p & -i|e_1\rangle_m|e_0\rangle_n|0\rangle_p & -i|e_1\rangle_m|e_0\rangle_n|0\rangle_p & -i|e_1\rangle_m|e_0\rangle_n|0\rangle_p & -|g\rangle_m|e_0\rangle_n|1\rangle_p \\
|e_0\rangle_m|g\rangle_n|1\rangle_p \Rightarrow & |e_0\rangle_m|g\rangle_n|1\rangle_p \Rightarrow & -i|e_0\rangle_m|e_1\rangle_n|0\rangle_p \Rightarrow & -|e_0\rangle_m|g\rangle_n|1\rangle_p \Rightarrow & -|e_0\rangle_m|g\rangle_n|1\rangle_p \\
|e_0\rangle_m|e_0\rangle_n|1\rangle_p & |e_0\rangle_m|e_0\rangle_n|1\rangle_p & |e_0\rangle_m|e_0\rangle_n|1\rangle_p & |e_0\rangle_m|e_0\rangle_n|1\rangle_p & |e_0\rangle_m|e_0\rangle_n|1\rangle_p \\
\end{array}
\quad \text{(EQ 20)}
$$

To perform a rotation through the third state in the 2-state model the simulator iterates over all the qubit states and performs the appropriate pair of rotations for each state. Instead of using a third state for each bit, the simulator uses a third state for each bit sequence, i.e. one state for each element in the complex vector space. This reduces the storage and computations complexity from $3^M$, for M qubits, to $2*2^M$.

Because the two rotations of a pair are performed at once we must combine the two error angles of the original pair to form a single error angle. If the pair of rotations are performed in succession, the combined error angle is just the sum of the two error angles. But if there is another $2\pi$ laser pulse performed between the pair, the combined angle now is the difference of the two angles. For example in Equation 20 the first $\tilde{U}_m(\pi, 0)$ laser pulse transforms the state $|g\rangle_m|g\rangle_n|1\rangle$ to $-i|e_0\rangle|g\rangle_n|0\rangle$. If this first laser pulse is performed perfectly no amplitude remains in the $|g\rangle_m|g\rangle_n|1\rangle$ state and the two $\tilde{U}_n(\pi, 0)$ transformations have no effect on the $|e_0\rangle|g\rangle_n|0\rangle$ state. However if the $\tilde{U}_m(\pi, 0)$ transformation is not performed perfectly, amplitude will remain in the $|g\rangle_m|g\rangle_n|1\rangle$ state after the laser pulse. The two $\tilde{U}_n(\pi, 0)$ laser pulses negate the sign of this state thereby also negating the sign of the error angle in the second $\tilde{U}_m(\pi, 0)$ laser pulse.

For bias operational errors subtracting the two error angles results in a cancellation effect. For noise, i.e. gaussian errors, the chance that the two error angles have the same sign is the same as the chance that they have opposite signs. Therefore in the average case subtracting the two angles is very similar to adding the two angles. We see this cancelling effect in the results shown in Section 4.0, where simulations with bias errors exhibit higher fidelities than do simulations with the same level of noise.

The cancellation effect illustrated above is a result of rotations within a gate, and is referred to as *intra* gate cancellation. This intra gate cancellation effect does not occur for all states because the intermediate $\tilde{U}_m(\pi, 0)$ transformations do not always negate the sign of the error states. However a cancellation effect can also occur between the transformations of separate gates. This *inter* gate cancellation occurs frequently for gates like the controlled-not gate where all the laser pulses are π pulses. For other types of gates, i.e. like those used in Grover's algorithm, there is not always a cancellation effect. This leads to the definition of the following two methods for computing the combined error angle:

- **Simple**: The combined error angle is computed as the difference of the two error angles from the pair of transformations. This models intra and inter gate cancellation for all gates. This is the method used in all benchmarks unless otherwise noted.

- **Mixed**: Use the simple model for all logic gates, i.e. controlled-not gates. For all other gates subtract the error angles if there is intra gate cancellation, and add the angles otherwise.

When we introduce operational errors into the transformations some amplitude may remain in the original states after the first transformation of a pair. The 2-State model cannot model the effect of subsequent transformations on this amplitude. The 2-State model also lumps together the third state error accumulation of multiple qubits. Comparing the results of the 3-State and 2-State models will determine the significance of these simplifying assumptions.

### 3.4.2 Reducing the complexity of the factor-15 problem

We can reduce the number of bits of *A* required in the factor-15 problem by observing that the period is determined in the *f(A)* circuit after performing at most three multiplications. This is because the period calculated by the *f(A)* circuit is a power of two, i.e. four. Shor suggests using $2L + 1$ qubits, to represent A, for the factorization of a number of size $2^L$. This increases the resolution of the FFT, and increases the probability of measuring a correct answer[Shor94][Joza96]. But because the period is a power of two there is no round off error and therefore no need to use all nine qubits.

Because we want to use the factor-15 problem to predict the behavior of larger factoring circuits, we use the number of multiplications suggested by Shor. Instead we use the special property of the factor-15 problem to reduce its complexity. This reduced model reuses the third A qubit for the last six multiplications. To reuse the qubit, we perform a rotation to remove the superposition of the qubit between the $|g\rangle$ and $|e_0\rangle$ states. We then clear all remaining amplitude in the $|e_0\rangle$ state and renormalize the state by increasing the amplitude of only those states which have amplitude due to error. This renormalization averages the error that has accumulated in both the $|g\rangle$ and $|e_0\rangle$ states.

## 4.0 Simulation results

In this section we present simulations which compare the accuracy of each of the reduced simulation techniques to the more detailed models. The *fidelity,* as defined in Equation 21, measures how close a state with error in it is to the correct result. The fidelity is defined as the inner product between the simulation with errors ($\psi$) and the correct result ($\varphi$). If there are no errors in the simulation, the fidelity will be equal to one, and if the errors cause the simulation to be totally orthogonal to the correct state, the fidelity will be zero. All fidelities are calculated after the *f(A)* circuit and before the FFT.

$$Fidelity = \|\langle\varphi|\psi\rangle\|^2 \qquad \text{(EQ 21)}$$

Table 3 shows all the benchmarks used in our simulation studies. Simulation using the 3-State model is only feasible for the three smallest benchmarks.

**TABLE 3. Benchmarks used in simulation studies. Ordered by increasing amount of complexity.**

| Benchmark | Number of qubits | Number of laser pulses | Description |
|---|---|---|---|
| f21 | 11 | 2,488 | Factor 21 using table lookup |
| grover | 13 | 1,838 | Circuit SAT using the Grover database search algorithm |
| mult | 16 | 8,854 | One modulo multiply step from the factor-15 problem |
| f15_3bit | 18 | 70,793 | Factor-15 problem using 3 qubits for A |
| f15_long | 24 | 70,904 | Factor-15 problem using all 9 qubits for A |

Table 4 shows the simulation time for each of the benchmarks simulating for operational errors. These simulation times assume a 300MHz Mega-Hertz DEC Alpha. For the larger problems we parallelized the simulations and ran them on a Cray T3E. This parallel version of the simulator achieves near linear speedup[ObDe98].

**TABLE 4. Simulation times for each benchmark**

| Benchmark | 2-State model (seconds) | 3-State model (seconds) |
|---|---|---|
| f21 | 3 | 125 |
| grover | 10 | 870 |
| mult | 288 | 88,614 |
| f15_3bit | 11,216 | N/A |
| f15_long | 193,248 | N/A |

## 4.1 Comparing the 2-State model to the 3-State model

Figure 7 through Figure 9 show the results of factoring simulations which compare the 3-State model to the 2-State model. In the simulation of the f21 benchmark, with only operational errors shown in Figure 7, the 2-State model is very accurate for errors with small values of $\mu$ and $\sigma$. The 2-State model is off by at most 0.066, for errors with a $\sigma$ of $\pi/128$ per gate. As discussed in Section 3.4.1, the accuracy of the 2-State model decreases with larger amounts of error because the 2-State model does not model second order errors in the third state. The fact that these second order effects come into play only for extreme amounts of error shows that the 2-State model is very accurate for realistic amounts of error.

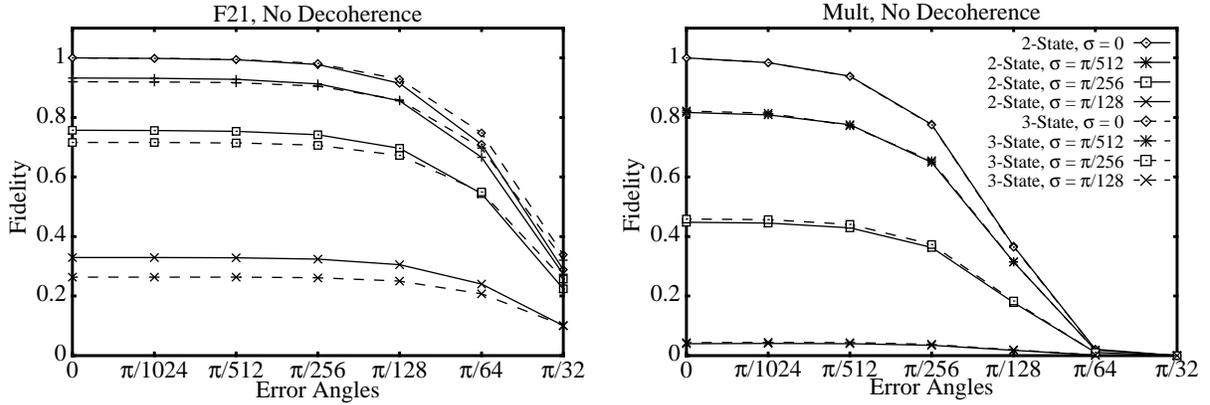

**FIGURE 7. Operational errors in the f21 and mult benchmark using the 3-State and 2-State models**

Figure 7 also shows that the 2-State model is even more accurate for the mult benchmark. The mult benchmark is about four times longer than the f21 benchmark, and therefore the total inaccuracy is amortized across a larger number of operations. This also shows that the inaccuracy of the 2-State model does not increase for larger simulations.

Figure 8 and Figure 9 show the results of simulations which include decoherence as well as operational errors. The simulations use the decay method, as described in Section 3.1.2, to model decoherence. The results using the 2-State model are exactly the same as those using the 3-State model for simulations without any operational error. This is to be expected because both models model the phonon mode in exactly the same way, and because there is not any operational error no amplitude is ever left in the third state.

The results also show that the 2-State model is very accurate for simulations which include operational errors as well as decoherence. The difference in fidelity of the two models in this case is the same as the difference for operational error considered alone. The difference between the two models, for the mult benchmark shown in Figure 8, is so small that the lines in the graph almost completely overlap each other.

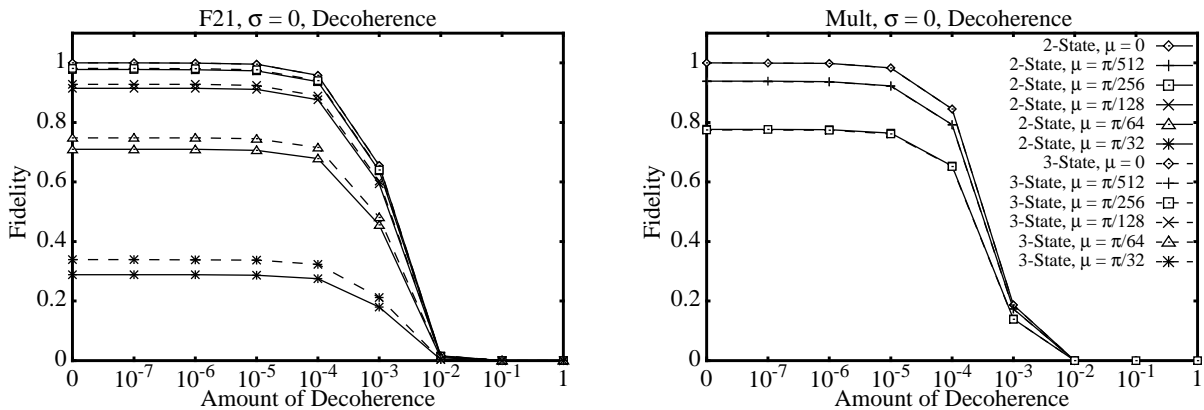

**FIGURE 8. Decoherence and Mean operational errors in the f21 and mult benchmarks using the 3-State and 2-State models**

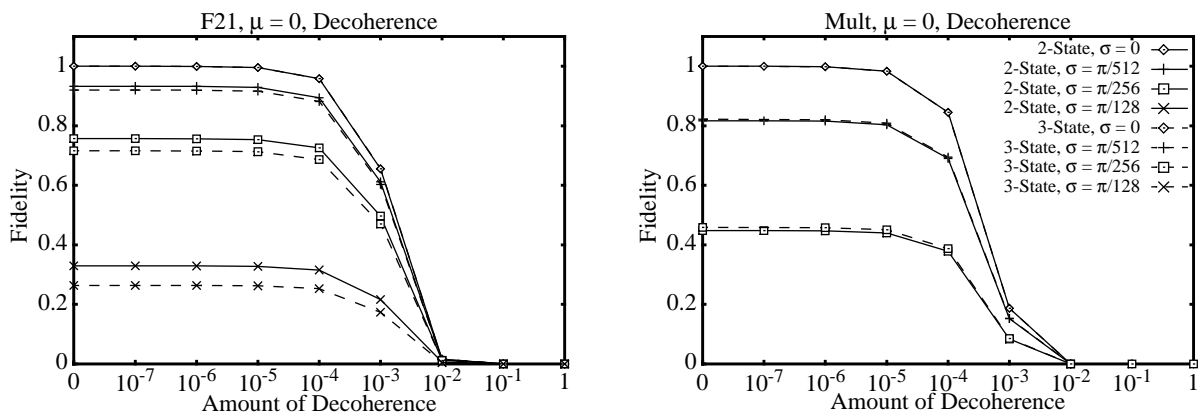

**FIGURE 9. Decoherence and noise in the f21 and mult benchmarks using the 3-State and 2-State models**

Figure 10 shows the results of simulations of operational errors for the grover benchmark. The first plot in Figure 10 uses the simple method to generate error angles, and the second figure uses the mixed method. With the simple method the 2-State model is accurate for errors with a mean less than $\pi/128$, but for larger errors the model is off by as much as 0.2.

Using the mixed method the 2-State model is very accurate. For error angles less than or equal to $\pi/128$ the 2-State model is indistinguishable from the 3-State model. The 2-State model is off by at most 0.02 for errors with a mean of $\pi/64$.

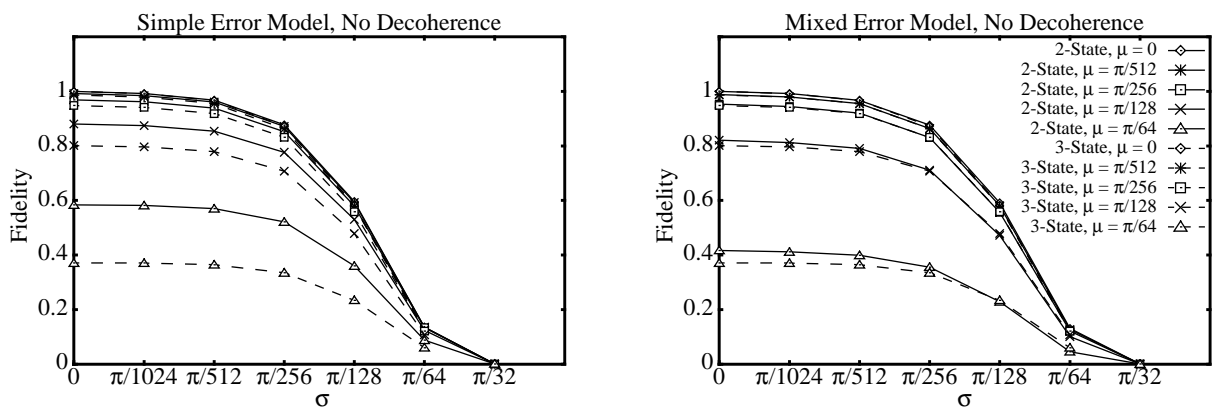

**FIGURE 10. Operational errors in the Grover database search benchmark using the 3-State and 2-State models**

Figure 11 shows 3-State and 2-State simulations of the grover benchmark with decoherence as well as operational errors. Operational error angles were computed using the mixed method. As both plots show the 2-State model is exact for decoherence without any operational error. As before the 2-State model accurately models operational errors as well, and the difference in fidelity between the two models is due solely to operational errors.

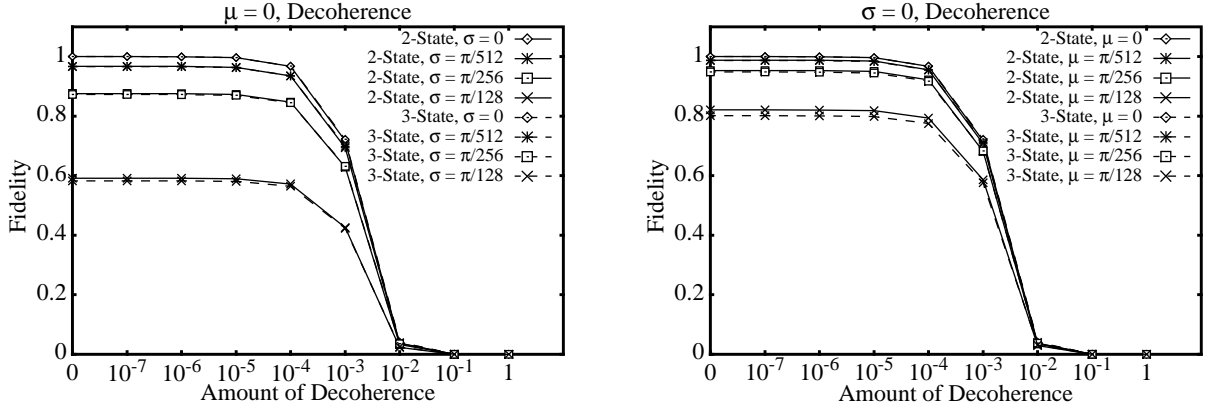

**FIGURE 11. Decoherence and operational errors in the Grover database search benchmark using the 3-State and 2-State models**

Table 5 summarizes the results, of the comparison of ending fidelities, from Figure 7 through Figure 11. It shows the maximum and average deviation of the 2-State model from the 3-State model. The grover benchmark, using the simple method to calculate the combined error angles, has the largest deviation. A maximum deviation of 0.2196 is clearly unacceptable. The 2-State model using the mixed method has a much more acceptable maximum deviation of 0.0461. For all other benchmarks and simulation models the maximum deviation is at most 0.0658. Also, for these cases, the average deviation is at most 0.0195 showing that the accuracy of the 2-State model in many cases is better than the maximum deviation.

**TABLE 5. Deviation between the 3-State and 2-State models**

| Benchmark and Simulation Model | Maximum Deviation | Average Deviation |
|---|---|---|
| f21, no decoherence | 0.0658 | 0.0195 |
| mult, no decoherence | 0.0116 | 0.0033 |
| f21, $\sigma = 0$, decoherence | 0.0504 | 0.0096 |
| mult, $\sigma = 0$, decoherence | 0.0019 | 0.0006 |
| f21, $\mu = 0$, decoherence | 0.0658 | 0.0116 |
| mult, $\mu = 0$, decoherence | 0.0108 | 0.0036 |
| grover, simple, no decoherence | 0.2196 | 0.0437 |
| grover, mixed, no decoherence | 0.0461 | 0.0078 |
| grover, mixed, $\mu = 0$, decoherence | 0.0087 | 0.0028 |
| grover, mixed, $\sigma = 0$, decoherence | 0.0197 | 0.0050 |

## 4.2 Accuracy of the f15_3bit benchmark

The f15_3bit benchmark is a simplification of the f15 benchmark which uses only three qubits for the A register. Because the f15 benchmark requires nine qubits for A, the f15_3bit benchmark is $2^6 = 64$ less complex. Figure 12 compares the two benchmarks for simulations of decoherence and operational errors. All lines for the f15 and f15_3bit benchmarks completely overlap each other showing that the f15_3bit benchmark accurately models the f15 benchmark.

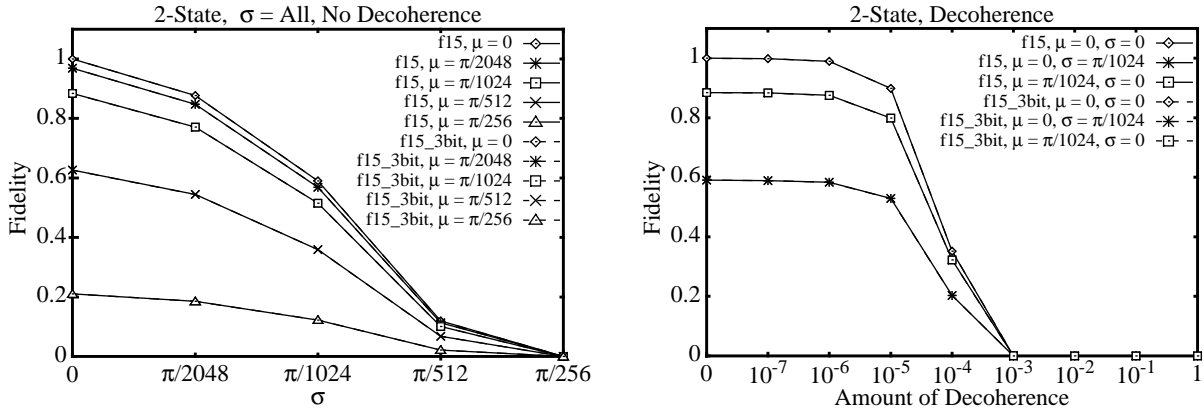

**FIGURE 12.** Comparison of the f15 benchmark to the f15_3bit benchmark for decoherence and operational errors

## 4.3 Modeling decoherence

In this section we consider both methods of modeling decoherence, spon_emit and decay, and show that they are essentially the same. We also show that there is very little correlation between decoherence and operational error. This allows us to simulate these two different types of errors separately, thereby reducing the total number of simulations which must be performed.

### 4.3.1 Comparing the different methods for modeling decoherence

Simulations using the spon_emit method explicitly include spontaneous emission. Multiple iterations are needed to average over the cases where emissions do and do not occur. The fidelity of the simulations where an emission occurs will be very low; whereas the fidelity of simulations without any emissions is very high. The average of both these cases is weighted based on the probability of emission.

For simulations using the decay method, the fidelity decreases over time. The rate at which it decays is based on the probability of emission, and therefore the fidelity at the end of the calculation is essentially the same as the fidelity obtained using the spon_emit method.

Figure 13 compares the fidelity of simulations using the decay and spon_emit methods for modeling decoherence. In the first plot we also introduce gaussian operational errors and in the second plot we introduce bias operational errors. As both plots show the fidelity computed by the spon_emit method is very close to the fidelity computed using the decay method. The biggest difference is 0.027 for a decoherence rate of $10^{-4}$, and no operational error. Also the difference between the two methods does not increase as we add more operational error.

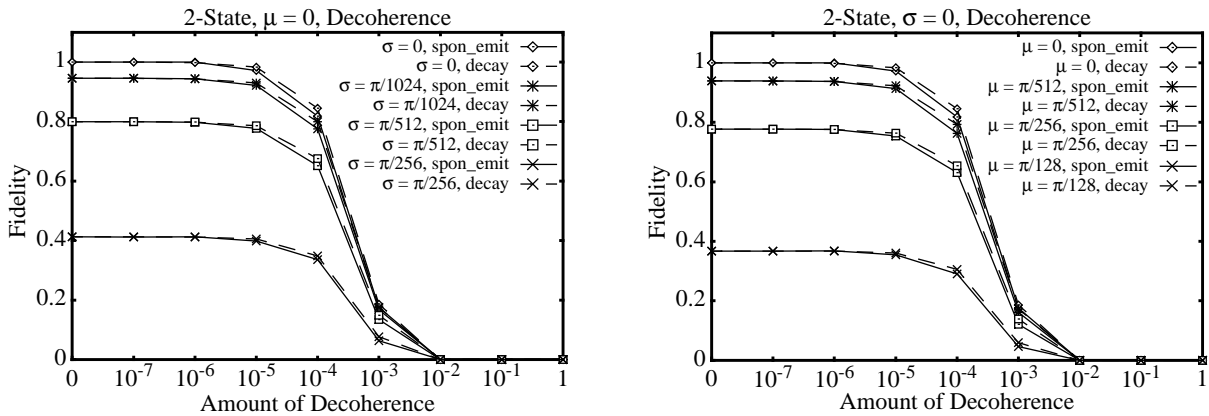

**FIGURE 13. Decoherence in the mult benchmark using both the spon_emit and decay methods**

Figure 14 compares the fidelity of the spon_emit and decay methods using the f15_3bit benchmark. As before the two methods produce very similar results. The greatest difference is for a decoherence rate of $10^{-4}$, where the decay method gives a fidelity which is 0.036 larger than the spon_emit method.

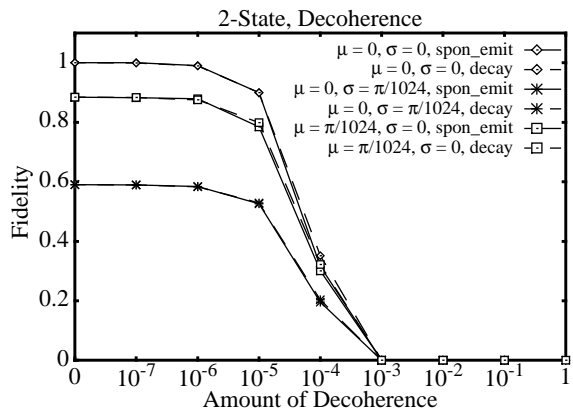

**FIGURE 14. Decoherence in the f15_3bit benchmark using both the spon_emit and decay methods**

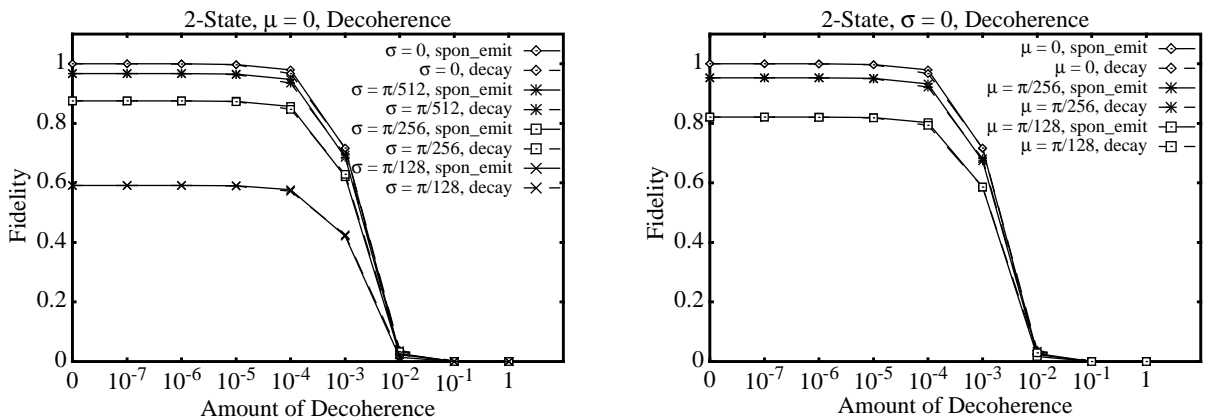

**FIGURE 15. Decoherence in the grover benchmark using both the spon_emit and decay methods**

The fidelity produced by the two methods is the closest for the grover benchmark, as shown in Figure 15. The spon_emit method calculates a fidelity which is at most 0.013 greater than the decay method, again for a decoherence rate of $10^{-4}$

### 4.3.2 The correlation between decoherence and operational errors

Both decoherence and operational error cause a degradation of the fidelity in a quantum computation. Decoherence degrades the fidelity through the decay of the phonon state, and operational error results in the accumulation of amplitude in unwanted states. The combined effect of these two factors is a degradation which is worse than either factor considered alone. We can represent the combined effect as:

$$F_{dec, op} = F_{dec} \bullet F_{op} + \Omega(F_{dec}, F_{op}) \tag{EQ 22}$$

Where $F_{dec}$ and $F_{op}$ are the fidelities of simulations for decoherence and operational error considered separately, and $\Omega(F_{dec}, F_{op})$ is the correlation between the two types of error. As Table 6 shows the correlation $\Omega(F_{dec}, F_{op})$ is very low. We calculated the correlation by running simulations which considered decoherence and operational error together. For all the benchmarks the maximum correlation is at most $1.14 \times 10^{-2}$. This result means that we can simulate decoherence and operational errors separately, and combine the results to obtain their collective effect on a calculation.

**TABLE 6. Correlation ($\Omega$) of decoherence and operational errors**

| Benchmark and Simulation Model | Maximum $\Omega$ | Average $\Omega$ |
|---|---|---|
| mult, $\mu=0$, $\sigma = \pi/1024 - \pi/64$ | $5.76 \times 10^{-5}$ | $3.63 \times 10^{-6}$ |
| mult, $\sigma = 0$, $\mu = \pi/1024 - \pi/64$ | $9.26 \times 10^{-3}$ | $5.51 \times 10^{-4}$ |
| f15_3bit, $\sigma = \pi/1024$, $\mu = 0$ | $4.15 \times 10^{-3}$ | $3.96 \times 10^{-4}$ |
| f15_3bit, $\sigma = 0$, $\mu = \pi/1024$ | $1.14 \times 10^{-2}$ | $8.76 \times 10^{-4}$ |
| grover, $\mu=0$, $\sigma = \pi/1024 - \pi/128$ | $1.78 \times 10^{-3}$ | $1.02 \times 10^{-4}$ |
| grover, $\sigma=0$, $\mu = \pi/1024 - \pi/128$ | $2.67 \times 10^{-3}$ | $2.36 \times 10^{-4}$ |

## 5.0  Conclusion

Simulating a quantum computer requires vast computational and processing resources due to the exponential nature of quantum mechanics. Simulating a detailed model like the Cirac and Zoller trapped ion scheme adds further to this complexity. In this paper we define a less complex model which accurately models the trapped ion quantum computer. This model represents an exponential decrease in complexity and allows us to simulate problems that would otherwise be infeasible.

In this paper we show simulations for the quantum factoring problem as well as for database search. The factoring benchmarks are implemented as sequences of one, two, and three input controlled-not gates. We perform simulations of the factor 21 problem using a table lookup method, and the factor 15 problem using repeated squaring. We also demonstrate a method for simulating the factor 15 problem whose complexity is lower than the original method by a factor of 64. Our implementation of the grover database search is the first such implementation for the trapped ion quantum computer. This implementation uses a sequence of rotations as well as controlled-not gates.

Through our simulations we also show that the two different types of error in a quantum computer, decoherence and inaccuracies, are uncorrelated. This allows us to perform separate simulations of the two different types of error and then combine the results instead of performing simulations of all possible combinations of the two types of error.

Simulation of the factor 15 problem for operational errors and decoherence would require about 30,000 years of simulation time if we were to use the most detailed simulation model and not use any of the simplifying techniques shown in this paper. This assumes that we simulate 25 different combinations of operational error, 8 different levels of decoherence and run four trials to average out the effects of random gaussian errors. Each of these simulations, using the 3-State model would take about 36 years. Performing the simulations at this detail is obviously not feasible, but with the methods described in this paper we can perform the same set of simulations in only 400 hours without an appreciable loss of accuracy. These methods will also allow us to simulate problems of increased size in the future.

## Acknowledgments

The authors are members of the Quantum Information and Computation (QUIC) consortium. We wish to thank our QUIC colleagues: Jeff Kimble, John Preskill, Hideo Mabuchi and Dave Vernooy. This work is supported in part by ARPA under contract number DAAH04-96-1-0386. Parallel simulations were performed on a Cray T3E at the Naval Oceanographic Office. The computer time was granted by the Challenge program from the DoD Modernization Office.